\documentclass[
    ,final            
  ]
  {aipproc}
\usepackage{graphicx}
\layoutstyle{6x9}
\begin{document}

\title{Loop quantization of spherically symmetric midi-superspaces : 
the interior problem}

\classification{4.60.Pp}
\keywords{Loop quantum gravity, spherical symmetry}

\author{Miguel Campiglia}{
address={
Department of Physics
The Pennsylvania State
University,\\ 104 Davey Lab, University Park PA 16802}}

\author{Rodolfo Gambini}{
address={
Instituto de F\'{\i}sica, Facultad de Ciencias, 
Igu\'a 4225, esq. Mataojo, Montevideo, Uruguay.}}

\author{Jorge Pullin}{
address={
2. Department of Physics and Astronomy, Louisiana State University,\\
Baton Rouge, LA 70803-4001}}

\begin{abstract}
We continue the study of spherically symmetric vacuum space-times in
loop quantum gravity by treating the interior of a black hole.  We
start from a midi-superspace approach, but a simple gauge fixing leads
to a Kantowski--Sachs form for the variables.  We show that one can
solve the quantum theory exactly in the (periodic) connection
representation, including the inner product. The evolution can be
solved exactly by de-parameterizing the theory and can be easily
interpreted as a semi-classical evolution plus quantum corrections. A
relational evolution can also be introduced in a precise manner,
suggesting what may happen in situations where it is not possible to
de-parameterize.  We show that the singularity is replaced by a bounce
at which quantum effects are important and that the extent of the
region at the bounce where one departs from classical 
general relativity depends on the initial data.
\end{abstract}

\maketitle

\section{Introduction}

In a previous paper \cite{cagapu} we have studied the quantization of
spherically symmetric midi-superspaces in loop quantum gravity.  We
were able to treat the space-time exterior to a black hole in both the
connection and the loop representation and recovered the quantization
that Kucha\v{r} \cite{kuchar} had constructed using the traditional
metric variables. The resulting quantum theories have wavefunctions
that are functions of the mass of the space-time that do not evolve in
time. The treatment being limited to the exterior, one could not probe
the more interesting issue of what happens to the singularity in the
quantum theory.  In this paper we would like to address that issue. We
will see that the same midi-superspace ansatz for the variables that
we used for the exterior can accommodate the interior of a black hole.
Through a simple gauge fixing one ends up with variables in
Kantowski--Sachs form.  We will construct the quantum theory in the
connection representation.  We will find that in the end one can
actually solve the theory exactly, and it is easy to describe the
evolution as a semi-classical one plus quantum corrections. We will
see that generically if one starts with a state that approximates
general relativity well at the cosmological horizon, quantum
corrections eventually become important and a ``bounce'' replaces the
cosmological singularity. The quantum solution therefore evolves past
the point where one had the classical singularity and generically
moves into another regime that approximates the classical theory well
and develops a cosmological horizon. The Hamiltonian constraint that
governs the evolution defines trajectories that depend on two constants
of the motion. Even though in the classical theory one of them may be
simply reinterpreted as a rescaling in the quantum regime both have
physical consequences. Generically, the bounce is not symmetric: the
extent of the regions where the quantum corrections are important is
not the same before and after the bounce.  However, the properties of
the ``inner'' cosmological horizon are the same as the one one started
with. If one were to naively match the resulting space-time to
Schwarzschild one would have that the horizon after the bounce would
have the same mass than the one of the horizon to the past of the
singularity.  We shall see that initial conditions can be set in such
a way that the evolution is symmetric with respect to the region where
the bounce occurs. The resulting symmetric wavefunctions will depend
only of arbitrary functions of one parameter, the mass of the Black
Hole.

\section{Classical setup}

We start with the same ansatz for the metric variables as in the 
exterior case. This is justified since the ansatz is good enough to
encompass the interior Schwarzschild solution written in Kantowski--Sachs
form. The kinematical setting for loop quantum
gravity in spherically symmetric situations is well established and
was discussed in detail by Bojowald and Swiderski \cite{boswi} and can
also be seen in our previous paper \cite{cagapu}.  There
is only one non-trivial spatial direction (the radial) which we call
$x$ since it is not necessarily parameterized by the usual radial
coordinate.  The canonical variables usual in loop quantum gravity are
a set of triads $E^a_i$ and $SO(3)$ connections $A_a^i$; after the
imposition of spherical symmetry one is left with three pairs of
canonical variables $(\eta,P^\eta,A_\varphi,E^\varphi,A_x,E^x)$ (they are
canonical up to factors involving the Immirzi parameter $\gamma$). 
The variables
$\theta$ and $\varphi$ are angles transverse to the radial direction
as in usual polar coordinates. Instead of using triads in the
transverse directions, one introduces a ``polar'' set of variables
$E^\varphi,\eta$ and their canonical momenta. The
relationship to more traditional metric variables is, 
\begin{eqnarray}
g_{xx}&=& \frac{(E^\varphi)^2}{|E^x|},\qquad g_{\theta\theta} = |E^x|,\\
K_{xx}&=&-{\rm sign}(E^x) \frac{E^\varphi}
{\sqrt{|E^x|}}\frac{(A_x+\eta')}{2\gamma},\qquad
K_{\theta\theta} = -\sqrt{|E^x|} \frac{A_\varphi}{2\gamma},
\end{eqnarray}
and the latter two are the components of the extrinsic
curvature. $\gamma$ is the Immirzi parameter of loop quantum gravity.

The Gauss law and diffeomorphism constraint read,
\begin{eqnarray}
  G&=& P^\eta+(E^x)'\\
  D&=& P^\eta \eta' + E^\varphi{A}_\varphi'-(E^x)' A_x.
\end{eqnarray}
and from now on we eliminate the $\eta$ variable and its conjugate
momentum by solving Gauss' law and introduce 
$K_\varphi={A}_\varphi/{2\gamma}$ and $K_x=(A_x+\eta')/2\gamma$. 
The diffeomorphism and Hamiltonian 
constraint now read,
\begin{eqnarray}
  D&=& E^\varphi K_\varphi'-(E^x)' K_x\\
H&=& 
-\frac{E^\varphi}{2\sqrt{|E^x|}} 
-2{\rm sg}(E^x){K_x {K}_\varphi \sqrt{|E^x|}}
-\frac{{K}_\varphi^2 E^\varphi}{2\sqrt{|E^x|}}
+\frac{\left((E^x)'\right)^2}{8\sqrt{|E^x|}E^\varphi}\nonumber\\
&&-\frac{\sqrt{|E^x|} (E^x)' (E^\varphi)'}{2 (E^\varphi)^2}
+\frac{\sqrt{|E^x|}(E^x)''}{2 E^\varphi}.
\end{eqnarray}
This form of the Hamiltonian differs slightly from the one we used in
\cite{cagapu} in the presence of the sign of the triad in the second
term. In the exterior treatment of our previous paper the sign was
constant and could be omitted, but it will change value when going
from $K_x$ to $-K_x$ so we need to keep it here. We have also rescaled
the lapse by a factor of sign of $E^x$, so from now on $N$ will be the
ordinary lapse times the sign.  The system has two pairs of canonical
variables $(K_x,E^x)$ and $(K_\varphi, E^\varphi)$ (we are taking
Newton's constant $G=1$).  The constraints are first class.

We now fix a gauge $\phi= (E^x)'=0$.
Preservation of $\phi=0$ in time partially determines one of the Lagrange
multipliers, the lapse,
\begin{eqnarray}
\dot{\phi}=0 &\Rightarrow& \{(E^x)',H\}=0\\
&=& (2 N K_\varphi)'=0\Rightarrow N'=0, 
\end{eqnarray}
where we have used the diffeomorphism constraint, that together with $\phi=0$
implies that $K_\varphi'=0$. The latter is a secondary constraint that also has
to be preserved in time. The preservation implies that $(E^\varphi)'=0$,
which is also a secondary constraint that is automatically preserved.
Therefore all the variables are now independent of the radial coordinate
$x$. The secondary constraints $(E^\varphi)'=0$ and $K_\varphi'=0$ are
second class and have to be imposed strongly. 

One is left with only one constraint, the Hamiltonian constraint,
\begin{eqnarray}
H&=& 
-\int dx {N\left(\frac{E^\varphi}{2\sqrt{|E^x|}} 
+{\rm sg}(E^x)2{K_x {K}_\varphi \sqrt{|E^x|}}
+\frac{K_\varphi^2 E^\varphi}{2\sqrt{|E^x|}}\right)
}
\end{eqnarray}
In the above expression $N, K_\varphi, E^\varphi, E^x$ are only
function of the evolution parameter, which we call $\tau$ and are
independent of the radial coordinate $x$. If one chooses $K_x$
independent of the radial coordinate initially, one is left with a
system where all canonical variables are functions of the evolution
parameter, which we will call $\tau$, only.  The integral on $x$ that
appears in the constraint is over a finite interval $L_0$ \cite{asbo}.

It is convenient to introduce variables that are more commonly used
in loop quantum cosmology. The new variables
are $b,c,p_b,p_c$ which are canonical up to a factor of the Immirzi
parameter and are defined as $b=\gamma K_\varphi$, $c=\gamma K_x L_0$
$p_b=E^\varphi L_0$, $p_c=E^x$. With these variables, the Hamiltonian
constraint has the form,
\begin{equation}\label{hamiltonian}
H = -\frac{p_b}{2} -\frac{2 b\, c\, p_c}{\gamma^2}-\frac{1}{2}\frac{b^2\,
p_b}{\gamma^2}
\end{equation}
where we have rescaled the constraint eliminating an overall factor
$1/\sqrt{p_c}$. Notice that a rescaling of the $x$ coordinate
also rescales $L_0$. This implies that a rescaling of $x$ does not
affect the value of $p_b$ and $c$. However, a change in $L_0$ will
rescale the value of $p_b$ and $c$. Therefore one can consider this
dependence as an additional gauge freedom of the cosmological 
counterpart of the Schwarzschild solution. The
evolution equations read,
\begin{eqnarray}
\dot{b} &=& 
-\frac{\gamma \bar{N}}{2} \left(1+\frac{b^2}{\gamma^2}\right),\label{11}\\
\dot{c} &=& -\frac{2\bar{N}\,b\,c}{\gamma},\label{12}\\
\dot{p}_b &=& \frac{\bar{N}}{\gamma}\left(2 c\,p_c+b\, p_b\right),\label{13}\\
\dot{p}_c &=& \frac{2\bar{N}\,b\,p_c}{\gamma},\label{14}
\end{eqnarray}
with $\bar{N}$ the rescaled lapse.

The usual approach is to choose one of the variables as clock, for
instance $p_c = \epsilon t(\tau)^2$ where $\epsilon=\pm 1$ since
an analysis of (\ref{11}-\ref{14}) shows that $p_c$ has a definite
sign throughout the evolution. The classical evolution contains
a singularity.
 Equation (\ref{14}) allows to determine the lapse, 
\begin{equation}
\bar{N} = \frac{\dot{t}(\tau) \gamma}{b\, t(\tau)}.
\end{equation}
Substituting in (\ref{11}) one gets an ordinary differential equation for $b^2$
that can be immediately integrated, leading to,
\begin{equation}
b^2 = -\gamma^2 +\frac{2\gamma^2 M}{t(\tau)},
\end{equation}
where $M$ is an integration constant. At the horizon one has that $b=0$ for
the horizon to be isolated, so $t=2M$ there. From here we get,
\begin{equation}
b=\epsilon \gamma \sqrt{-1 +\frac{2 M}{t(\tau)}},
\end{equation}
where we have chosen the sign such that the original (un-rescaled)
lapse remains positive when one tunnels through the singularity.

Solving now (\ref{12})  we get,
\begin{equation}
c = \frac{\gamma M \epsilon C_0}{t(\tau)^2}
\end{equation}
where $C_0$ is a constant. Using equation (\ref{13}) and the Hamiltonian
constraint which implies that $p_b=0$ when $b=0$, we get,
\begin{equation}
p_b=2 C_0 \sqrt{t(\tau)\left(2M-t(\tau)\right)}.
\end{equation}
The constant $C_0$ can be chosen by rescaling $L_0$, so 
we do so in such a way that $2 C_0=1$. If we choose the parameterization
$t=\tau$ the lapse is completely determined and takes the form,
\begin{equation}
N = \frac{1}{\sqrt{\frac{2M}{t}-1}}.
\end{equation}

This leaves us with the traditional line element for Kantowski--Sachs,
\begin{equation}
ds^2 = -\left(\frac{2M}{t} -1\right)^{-1} dt^2 
+\left(\frac{2M}{t} -1\right)dx^2
+t^2 d\Omega^2.
\end{equation}

We therefore see that we recover the traditional Kantowski--Sachs solution
within the midisuperspace we started from.  The system has
only one constant of integration corresponding to one mechanical 
degree of freedom given by the value of $M$, just like in the exterior
case. The slicing corresponding to the $t={\rm const.}$ corresponds in 
the Kruskal diagram to hyperboloids $r={\rm const.}$.

\section{The quantum theory}
\subsection{Quantization}

We can now proceed to ``holonomize'' the classical expressions found
in order to carry out a loop quantization. The dynamical variables in
the Hamiltonian are $b, p_b$ $c$ and $p_c$.  Introducing a Bohr
compactification for $b$ and $c$ one obtains a ``loop quantum
cosmology'' which has been discussed by Ashtekar and
Bojowald \cite{asbo} and Modesto \cite{modesto}. Instead of repeating
their construction we will limit ourselves to introducing a technique
that will be useful to discuss with more detail the issue of the bounce. 
More recently, Boehmer and Vandersloot \cite{vandersloot} carried out 
a study of the holonomized semiclassical theory both with the $\mu$
and $\bar{\mu}$ approaches. Their results for the $\mu$ case parallel the
ones we discuss in the next subsection, although our approach is relational,
whereas they  work in a parameterized way.

The Bohr compactification implies that the configuration variables
take values in a compact space. The kinematical Hilbert space of wavefuctions
is therefore given by periodic functions of the configuration variables.
The ``holonomized'' version of the Hamiltonian constraint (\ref{hamiltonian})
is an expression that is well defined as a quantum operator acting on such
a Hilbert space. It therefore involves the configuration variables in 
periodic fashion. The expression depends on a parameter $\mu$
and reproduces the Hamiltonian (\ref{hamiltonian}) in the limit $\mu\to 0$.
It takes the form,
\begin{equation}
H = -\frac{p_b}{2} -\frac{2 \sin(\mu\,b)p_c\sin(\mu\,c)}{\mu^2 \gamma^2}
-\frac{p_b\sin(\mu\,b)^2}{2 \mu^2 \gamma^2}.\label{Hhol}
\end{equation}
Upon quantization in the Bohr compactified space we keep $\mu$ finite,
as is customary in loop quantum cosmology. The use of a fixed value of
$\mu$ breaks the scale invariance of the Hamiltonian as it can be
easily seen from (\ref{Hhol}).

In the Hilbert space considered  the momentum
variables are,
\begin{eqnarray}
\hat{p}_b \Psi(b,c) &=& i \gamma \ell_{\rm Planck}^2 
\frac{\partial \Psi(b,c)}{\partial b},\\
\hat{p}_c \Psi(b,c) &=& i \gamma \ell_{\rm Planck}^2 
\frac{\partial \Psi(b,c)}{\partial c},
\end{eqnarray}
and on this space $\hat{b}$ and $\hat{c}$ are not well defined but
periodic functions of them are.

In order to quantize, it is better to rearrange the classical Hamiltonian
constraint in a way that it is easy to ``deparameterize'',

\begin{equation}\label{25}
-p_c = H_{\rm True} = \frac{1}{4} \frac{\mu^2 \gamma^2 + \sin(\mu\,b)^2}
{\sin(\mu\,b)\sin(\mu\,c)}p_b,
\end{equation}
which upon quantization yields the following form for the symmetric
form of the ``true'' Hamiltonian,
\begin{eqnarray}\label{26}
\hat{H}_{\rm True}\Psi(b,c)&=& -\frac{i \ell_{\rm Planck}^2 \gamma}
{4\sin(\mu\,c)\sin(\mu\,b)} \left[
\left(\sin(\mu\,b)^2+\mu^2\gamma^2\right)
\frac{\partial \Psi(b,c)}
{\partial b}\right.\\ 
&&+\left.\frac{\mu\cos(\mu\,b) }{2\sin(\mu\,b)}
\left(\sin(\mu\,b)^2 -\mu^2 \gamma^2\right)
\Psi(b,c)\right].\nonumber
\end{eqnarray}


Equations (\ref{25},\ref{26}) can be solved using standard techniques
for linear partial differential equations that allow to reduce the
system to a system of ordinary differential equations. The solution 
can be written in the form,
\begin{equation}\label{27}
\Psi\left(b,c\right) = 
A\left(b,c\right) \exp\left( \frac{ik}{l_P^2}
S\left(b,c\right)\right),
\end{equation} 
with $k$ a constant with dimensions of length squared, 
and where $A$ is given by,
\begin{equation}\label{abc}
A(b,c) = C_1\sqrt{\frac{\sin^2(\mu\,b)+\gamma^2\mu^2}{\left(1+\gamma^2\mu^2
\right)|\sin(\mu\,b)|}}
\end{equation}
where $C_1$ is a normalization constant and $S(b,c)$ is a solution of 
\begin{equation}
\left[\frac{\mu^2\gamma^2}{4\sin(\mu\,b)\sin(\mu\,c)}
+\frac{\sin(\mu\, b)}{4\sin(\mu\, c)}\right]
\frac{\partial S(b,c)}{\partial b} +
\frac{\partial S(b,c)}{\partial c}
=0,
\end{equation}
whose general solution is an arbitrary function $S=\Phi(w(b,c))$ 
with $w(b,c)$ a constant of the motion given by 
\begin{equation}
w(b,c) = \ln\left(\frac{\sin(\mu\,c)}{\cos(\mu\,c)+1}\right)
+ 4 \frac{\tanh^{-1}\left(\frac{\cos(\mu\,b)}{\sqrt{\gamma^2\mu^2+1}}\right)}
{\sqrt{\gamma^2\mu^2+1}}.
\end{equation}

\subsection{Semiclassical approximation}

As in the usual eikonal approximation $S$ contains the classical
behavior of the solutions. Here, however, since we are working with
the holonomized theory this will include the effective semiclassical
dynamics of loop quantum gravity. This holonomized theory depends on
an arbitrary parameter $\mu$, as is usual in the minisuperspace
context.  If one wishes to identify the value of the parameter $\mu$
with similar concepts of the full theory, for instance, based on the
quanta of area, the value of $\mu$ depends on Planck's constant. That is how
one understands the classical holonomized theory as a semiclassical
approximation to the dynamics of the model. The analysis can be
done for any function $\Phi(w)$ and leads to solutions that differ by
constant scalings, which can be reabsorbed in the constant $k$. We
choose $\Phi(w)=w$.  Unlike the usual case in unconstrained systems,
where it depends on two, here it depends on only one integration
constant $k$.

With the solution for $S$ we can study the classical behavior of the
system. In the usual Hamilton--Jacobi theory one starts by computing
the canonical momenta of the original configuration variables, in this
case $b$ and $c$,


\begin{eqnarray}
\label{pb}
p_b &=& \frac{4 k \sin(\mu\,b) \mu}
{\gamma^2\mu^2 +1 -\cos(\mu\,b)^2}\\
p_c &=& -\frac{k \mu}{\sin(\mu\,c)}\label{pc}
\end{eqnarray}
and the canonical momenta of the constants of integration, in this case
only one, $k$, which are also constants of the motion,
\begin{equation}
p_k = -\ln\left(\frac{\sin(\mu\,c)}{\cos(\mu\,c)+1}\right)
-\frac{4 \tanh^{-1}\left(\frac{\cos(\mu\,b)}{\sqrt{\gamma^2\mu^2+1}}\right)}
{\sqrt{\gamma^2\mu^2+1}}
\end{equation}

We will now consider the initial conditions, which we will take at the
cosmological horizon. We know that there one has $b=0$ (isolated
horizon condition) and one identifies $p_c=4M^2$ so it can be
isometric to the usual expressions in Schwarzschild.  The initial
value of $c$ which we label $c_0$ is in principle arbitrary.
Substituting in the equations (\ref{pb},\ref{pc}) we get $p_b=0$
initially and the constant $k$ takes a value determined by $c_0$ and
$M$ and has dimensions of length squared. It is given by:
\begin{equation}
k=-\frac{4 M^2}{\mu} \sin(\mu c_0).
\end{equation}

 Finally, the equation for $p_k$,
which is independent of time, evaluated at different times, yields
an equation from which we can obtain one of the configuration 
variables in terms of the other and the initial conditions. In particular,
if we choose to solve for $b$, we get,


\begin{eqnarray}
b &=& \frac{1}{\mu} \cos^{-1}\left[\sqrt{1+\gamma^2\mu^2} \tanh\left(
\tanh^{-1}\left(\frac{1}{\sqrt{1+\gamma^2\mu^2}}\right)\right.\right.\nonumber\\
&&\left.\left. +
\frac{\sqrt{1+\gamma^2\mu^2}}{4}\left(-\ln\left(\tan(\frac{\mu\, c}{2})\right)+
\ln\left(\tan(\frac{\mu\,c_0}{2})\right)\right)\right)\right]
\end{eqnarray}

Substituting $b$ into the expressions for $p_b$ and $p_c$ we have all
canonical variables as functions of $c$ which operates as time
variable.  The analytic expressions are lengthy. To analyze their
behavior it is best to resort to numerical evaluation. For this we
need to choose numerical values for the various constants. For
simplicity we take the Immirzi parameter to be unity $\gamma=1$, which
is close to the value that stems from black hole entropy. The
parameter $\mu$ is arbitrary for the classical theory, although usual
loop quantum cosmology arguments lead to $\mu=\sqrt{3}/4$. However
this leads to a ``bounce'' happening far from the Planck radius. This
led in loop quantum cosmology to consider a variable value of $\mu$.
We will not pursue that route here. We will just consider $\mu$ as a
free (but small) parameter. In the numerical computations shown here
we have taken $\mu=0.02$.

The problem has two constants of the motion. One of them is the
initial value of $c=c_0$. The other is the value of the ``mass'' ---if
one is viewing the problem as a Schwarzschild interior---, which is
determined by the value of $k$. When the problem is being considered
as a Schwarzschild interior, it is natural to rescale $c_0$ by varying
$L_0$ as we noted before. In the holonomized theory, however, this
rescaling freedom is not present at the quantum level due to the
finite value of the ``area quantum'' $\mu$, and different values of
$c_0$ lead to different behaviors, in particular, at the bounce. We
shall analyze this effect and show that a preferred value is naturally
selected once $\mu$ is fixed.

An interesting check is to note that in the limit $\mu\rightarrow 0$
the semi-classical solution of the holonomized quantum theory recovers
the traditional Kantowski--Sachs space-time of general relativity, as
expected.  In particular one has that the independent constant $C_0$
of Kantowski-Sachs is related with the initial constant $c_0$ as
$C_0=4 c_0 M/\gamma$.

\begin{figure}[htbp]\label{fig1}
  \centerline{
\includegraphics[height=7cm]{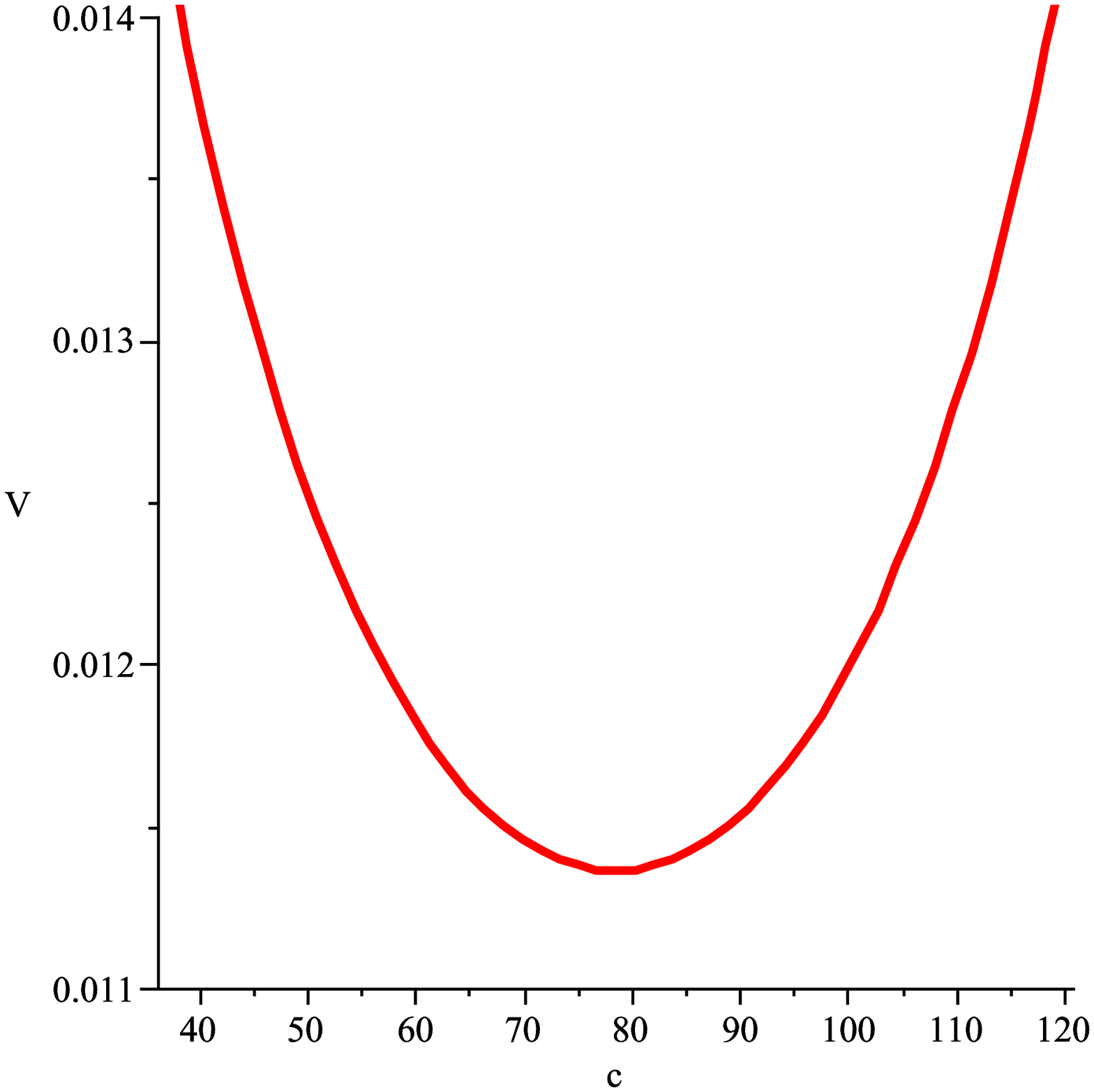}\,\,\,\,\,
\includegraphics[height=7cm]{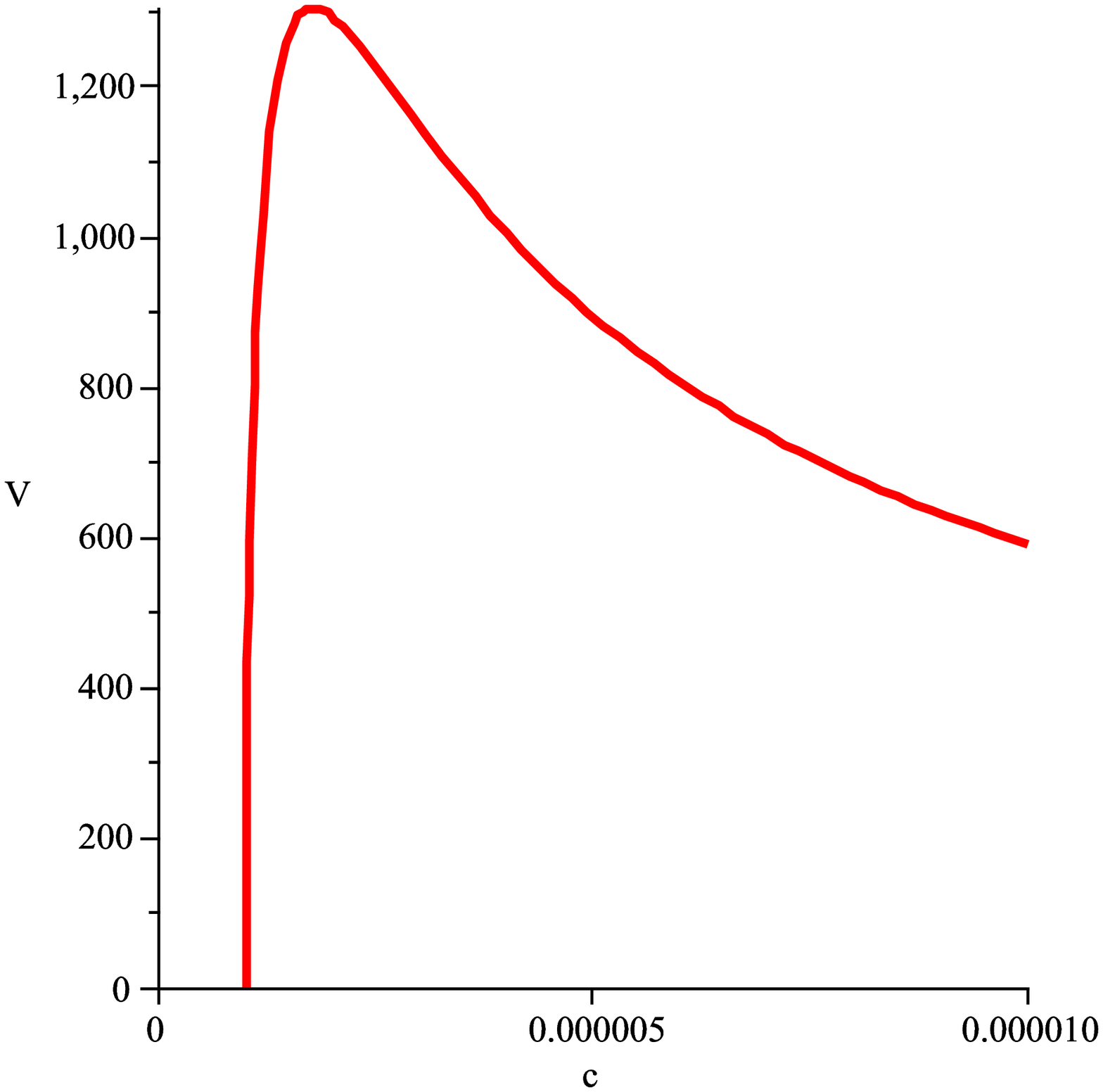}
}
\caption{
  The volume as a function of the time variable $c$. We see a bounce
  occurring in the place where usual Kantowski--Sachs would have a
  singularity (left panel). The right panel shows the volume as a
  function of time close to the Kantowski--Sachs horizon ($b=0$),
  where the volume vanishes. Calculations use $M=5\times10^5 M_{\rm Planck}$.}
\end{figure}

We have evaluated numerically the exact semi-classical solutions given
before in order to study their properties. In figure \ref{fig1} we
show the volume as a function of the time variable $c$.  One sees a
bounce occurring where the singularity would have appeared in general
relativity. The volume does not vanish and the co-triad does not
diverge. The volume goes to zero at the Kantowski--Sachs horizon,
which occurs at the past an future horizons where $b=0$. We have
chosen $b=0$ at the initial data, where we chose $c=c_0=10^{-6}$ (the
value $c_0=0$ yields an ill-defined evolution).  The volume then also
vanishes at the final point, at $c=\pi/\mu-c_0$, irrespective of the
initial value of $c_0$.

\begin{figure}[htbp]\label{fig2}
  \centerline{
\includegraphics[height=7cm]{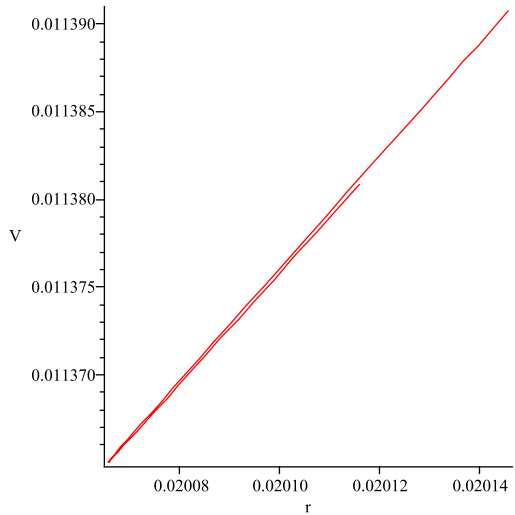}
}
\caption{
  The volume as a function of the ``radial'' time variable $r^2=p_c$.
  It can be seen that it contracts till it achieves a minimum value,
  it then expands at the same expansion rate as it contracted. This is
  achieved if one chooses the initial data for $c_0$ at the horizon in
  such a way that the evolution is symmetric at the ``bounce''. This,
  in particular implies that the region where the holonomized theory
  departs from general relativity is minimized. For other choices, the
  region of departure is larger. }
\end{figure}

In figure \ref{fig2} we show the volume as a function of the
``radial'' variable $r^2=p_c$. The volume is multi-valued and it
corresponds to the contracting and expanding phases. The rate at which
the volume contracts depends on the value of $c_0$. We have chosen it
here in such a way that the contracting and expanding rates are very
close to each other. This, however, is a choice. In particular, the
region where the holonomized theory departs from general relativity
will change depending on such choice. If one chooses, as we did, a
``symmetric'' evolution, the region of departure is confined near the
bounce in a symmetric way. Other choices can lead to the region
extending far away from the bounce either into the future or the past
of it. It is remarkable that no matter what choice of $c_0$ initially,
the system will eventually go past the bounce and return to a regime
where the holonomized theory approximates general relativity and a
future horizon will develop with the same mass value. That is, the
presence of both horizons and the bounce is robust. How the bounce
occurs and when the holonomized theory starts and stops its departure
from general relativity depends on the initial data. This is a
reflection of the instabilities that were noticed in the recursion
relations of the loop representation treatment of these models
\cite{khanna}.

\begin{figure}[htbp]\label{fig3}
  \centerline{
\includegraphics[height=7cm]{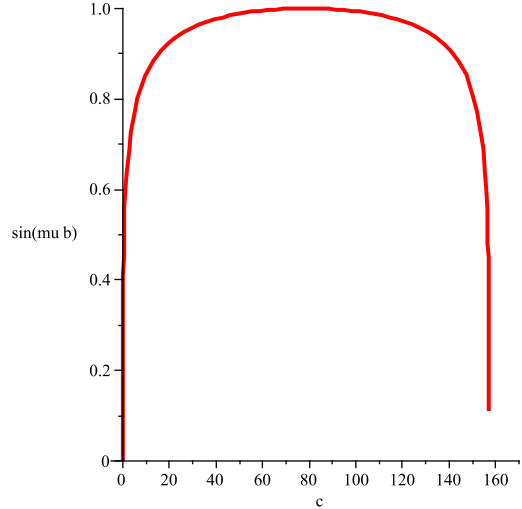}
}
\caption{
  The sine of $\mu b$ vs $c$. When the sine is close to one the
  holonomized theory departs from general relativity significantly. We
  see that both theories agree at the initial and final moments, when
  the cosmological horizons are present. The diagram is symmetric in
  spite of its appearance, it is hard for the plotting program to
  numerically get to the points near the final horizon, since the
  points gather exponentially.}
\end{figure}

Figure \ref{fig3} shows the sine of $\mu b$ versus the time variable
$c$.  The sine of $\mu\,b$ divided by $\mu$ is the variable in the
holonomized theory that replaces the connection $A_\phi$ in the usual
theory. Both theories agree when the sine is small.  We see that
indeed the sine is small close to the horizons and maximal at the
bounce. The time variable $c$ chosen is such that the region close to
the ``bounce'' measured in terms of the radial variable is achieved
very fast and lasts for a long ``time'', therefore the sine of $\mu b$
is large through most of the evolution. One can see that when measured
in terms of the volume the connection departs from it classical value
close to bounce.
  
\subsection{Quantum corrections}

We have discussed the classical approximation. Let us now proceed to
consider the quantum corrections, which involve the prefactor $A(b,c)$
in equation (\ref{27}). We start by noting that the proposed solution
for the wavefunction satisfies the boundary conditions implied by the
kinematical Hilbert space.  That is,
$\Psi(-\pi/\mu,c)=\Psi(\pi/\mu,c)$,
$\Psi(b,-\pi/\mu)=\Psi(b,\pi/\mu)$.  We
have required that the boundary conditions be periodic in $c$ and $b$
with period $2\pi/\mu$.  It turns out that the only region of phase
space of importance is between $0$ and $\pi/\mu$.  In the other three
regions one obtains the same $|\Psi|^2$ and the physical predictions
are the same. It might appear surprising to impose boundary conditions
in the variable $c$, which we are using as ``time''.  In principle one
would think that specifying ``future'' boundary values would violate
causality.  Remarkably, this is not the case. The value of $c_0$ is
arbitrary and the final result of the evolution is in no way
determined by the periodicity imposed.

The dynamical Hilbert space is therefore given by arbitrary functions
of the two constants of the motion $c_0,M$. As the Schwarzschild
external solution depends only of $M$ one can restrict the Hilbert
space to the symmetric semi-classical solution around the bounce that
are the ones whose behavior departs from the classical only close to
radial distances of the order of Planck scale. Given $\mu$ this is
equivalent to choose $c_0$ such that
\begin{equation}
{c_0}=\frac{1}{\mu}\tan^{-1}\left[\frac{2\Theta^{2 \Delta^{-1}}}
{\Theta^{4\Delta^{-1}}-1}\right],
\end{equation}
with
\begin{equation}
\Theta=\frac{
{2}^{-\Delta/4} \left( 2+\sqrt {2}
 \right) ^{\Delta/2}}
{\left(\sqrt{2}+2\Delta\right)\gamma^2\mu^2}
\left(\left[-\sqrt {2}+4+2\,
\Delta\right]\gamma^2\mu^2
+\left(4-2\,\sqrt{2}\right)\left(1+\Delta\right)\right)
) 
\end{equation}
and $\Delta=\sqrt{1+\gamma^2\mu^2}$. Going to the full quantum
theory, however, will not allow to choose precisely the values of 
$c_0,M$ that correspond to the symmetric bounce, and one will have
a spread implying that the region where the bounce occurs will be
larger than the one we consider here in the symmetric case.


Finally, one can show that the evolution in $c$ is unitary. In fact
one can see that $H_{\rm True}$ is self-adjoint by using the following
technique \cite{porto4}: Check the dimensionality of the two
subspaces: ${\cal K}_+ = {\rm ker}( H_{\rm True}+i)$ and ${\cal K}_- =
{\rm ker}( H_{\rm True} - i)$ They are self-adjoint if and only if
both spaces have the same dimensionality.  In this case the
wavefunctions $\Psi$ multiplied by the exponential factor
\begin{equation} 
\exp\left(\mp4\, \tanh^{-1} \left( {\frac {
\cos \left( \mu\,{\it b}\right)  \sin \left( \mu\,{\it c} \right)}
{\gamma \mu \left({{\mu}^{2}{\gamma}^{2}+1}\right)}}\right)\right)
\end{equation} 
respectively belong to each kernel and are normalizable in the
kinematical space. An important observation is in order here, the
elements of the physical space $\Psi(b,c)$ are also normalizable
with the inner product of the kinematical space and therefore they can
be also used to define a relational evolution in terms of conditional
probabilities, as for instance in the proposal of Page and Wootters
\cite{PaWo,njp}. A detailed study of these definition of the evolution
that do not require a deparametrization of the constraint will be
given elsewhere.

We have therefore carried out a quantization of the interior of the
Schwarzschild space-time using loop quantum gravity techniques.  We
have shown how the singularity is replaced by a bounce, and what
conditions are needed for the bounce to occur in a regime of Planck
energy. We have also outlined how one would construct the quantum
theory for the model. It is interesting to compare the interior and
exterior treatments. In the latter (and also in the complete space-time
treatment of Kucha\v{r} \cite{kuchar}) one is left with an quantum theory
in which wavefunctions are arbitrary functions mass of the space-time,
which is conserved. Here, while treating the interior as a cosmology,
we are left with arbitrary functions of an observable, which evaluated
on the initial data is determined entirely by $c_0$. This is the variable
conjugate to $p_c$, whose initial value is associated with
the mass of the Schwarzschild space-time. Therefore the two pictures
are clearly reconciled.

This work was supported in part by grant NSF-PHY-0554793, 9907949,
0456913, funds of the Hearne Institute for Theoretical Physics, The
Eberly research funds of PennState, FQXi, CCT-LSU and Pedeciba (Uruguay).

\end{document}